\documentclass[sigconf]{acmart}

\DeclareUnicodeCharacter{FF0C}{,}

\AtBeginDocument{%
  \providecommand\BibTeX{{%
    \normalfont B\kern-0.5em{\scshape i\kern-0.25em b}\kern-0.8em\TeX}}}

\usepackage{booktabs}
\usepackage{multirow}
\usepackage{threeparttable}
\usepackage{comment}
\usepackage{subcaption}
\usepackage{xspace}

\newcommand\approach{QuMATL}
\newcommand\task{MATL}

\newcommand{\ie}{i.e.\@\xspace}

\begin{document}

\title{QuMATL: Query-based Multi-annotator Tendency Learning}



\author{Liyun Zhang}
\affiliation{%
  \institution{D3 Center, Osaka University}
  \country{Japan}
}
\email{zhang.liyun@ids.osaka-u.ac.jp}

\author{Zheng Lian}
\affiliation{%
  \institution{Institute of automation, Chinese academy of science}
  \country{China}
}

\author{Hong Liu}
\affiliation{%
  \institution{Xiamen University}
  \country{China}
}

\author{Takanori Takebe}
\affiliation{%
  \institution{Cincinnati Children's Hospital Medical Center}
  \country{Japan}
}

\author{Yuta Nakashima}
\affiliation{%
  \institution{D3 Center, Osaka University}
  \country{Japan}
}



\begin{abstract}
  Different annotators often assign different labels to the same sample due to backgrounds or preferences, and such labeling patterns are referred to as ``tendency''. In multi-annotator scenarios, we introduce a novel task called Multi-annotator Tendency Learning (\task), which aims to capture each annotator's tendency. Unlike traditional tasks that prioritize consensus-oriented learning, which averages out annotator differences and leads to tendency information loss, \task\ emphasizes learning each annotator's tendency, better preserves tendency information. To this end, we propose an efficient baseline method, Query-based Multi-annotator Tendency Learning (\approach), which uses lightweight query to represent each annotator for tendency modeling. It saves the costs of building separate conventional models for each annotator, leverages shared learnable queries to capture inter-annotator correlations as an additional hidden supervisory signal to enhance modeling performance. Meanwhile, we provide a new metric, Difference of Inter-annotator Consistency (DIC), to evaluate how effectively models preserve annotators' tendency information. Additionally, we contribute two large-scale datasets, STREET and AMER, providing averages of 4,300 and 3,118 per-annotator labels, respectively. Extensive experiments verified the effectiveness of our \approach. 
\end{abstract}












\maketitle

\section{Introduction}
\label{sec:intro}

\begin{figure}[t]
  \centering
  \includegraphics[width=\linewidth]{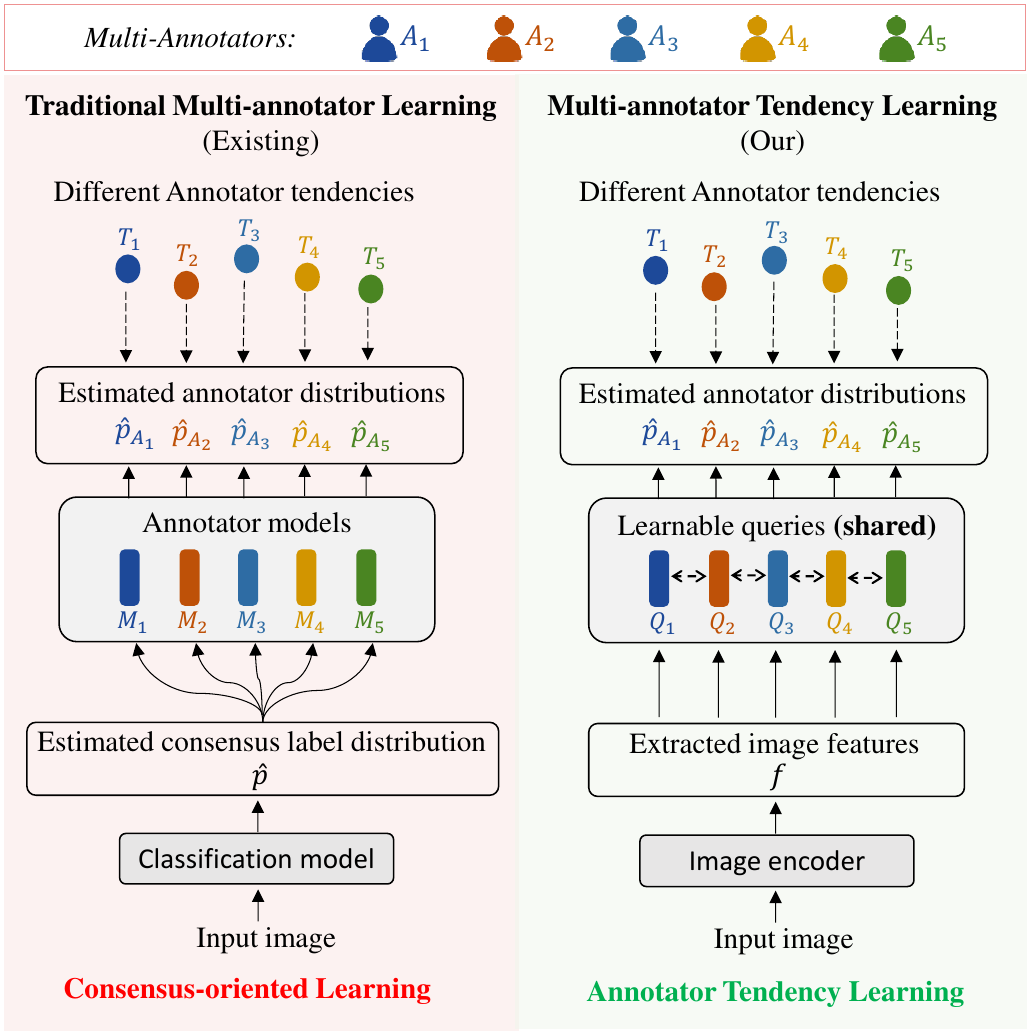}
  \caption{Task and model comparisons. We take five different annotators as an example. They give different labels to the same image due to preferences, we define such labeling patterns as \emph{tendency}, corresponding to circles at different positions reflected dataset labels. The existing task in left focus on a consensus-oriented learning, uses multi-annotator supervision to indirectly train for a single consensus label prediction, easily tends to average out annotator differences leading to tendency information loss. Our introduced new task \task\ focuses on learning each annotator's tendency by only building a dedicated annotator model, which can better preserve tendency information. Compared to existing methods, our proposed \approach\ uses lightweight query to represent each annotator for tendency modeling and save costs of building separate conventional models for each annotator, leverages shared learnable queries to capture inter-annotator correlations as an additional hidden supervisory signal to enhance modeling performance.}
  \label{overview}
\end{figure}

In real-world multi-annotation scenarios, such as medical image analysis \cite{PADL}, sentiment analysis \cite{lian2023mer}, and visual perception \cite{CNN-CM}, different annotators often provide different labels to the same sample \cite{LFC} due to different personal backgrounds, subjective interpretations, and preferences. We refer to such labeling patterns exhibited by each annotator as \emph{tendency}, and we introduce a new task called \emph{Multi-annotator Tendency Learning (\task)}, which aims to capture each annotator's tendency.

Let us suppose we have five annotators $A_i, i \in \{1, \cdots, 5\}$ for example, as shown in Figure~\ref{overview}, and use circles at different positions to represent the varying tendency $T_i$ of annotators reflected in the multi-annotator dataset labels. The left part represents the traditional multi-annotator learning tasks, which typically implement consensus-oriented learning. For example, it often builds a classification model that outputs a unique consensus label distribution prediction $\hat{p}$ for an input image and builds annotator models that capture different tendency of each annotator to predict each annotator's label distribution$\hat{p}_{A_i}$ the input image. These models are trained with a dataset with multi-annotator labels. This way, a single model is trained through the labels from multiple annotators. This consensus-oriented learning often has an impact on the effects of the annotator tendency capturing process, such as reducing different annotators' nuanced judgments on individual samples to simplified prediction-level biases, which easily causes tendency information loss during annotator modeling, \ie, the learned annotator models reduced differences among annotators' tendencies.

In contrast, the right part illustrates our introduced \task\ task, which focuses on annotator tendency learning. It emphasizes building models for different annotators to learn their tendencies, and then attempts to validate that these trained models with more tendency information effectively help implement specific tasks better. To this end, we proposed a query-based multi-annotator tendency learning (\approach) method. It assumes that differences between annotators are attributed to what observers focus on. Our method leverages learnable queries (${Q}_i$) of query transformer (Q-Former) \cite{BLIP2} to model different annotators' tendencies, uses cross-attention module to interact with image features $f$ extracted by image encoder from the input image to capture different annotators' nuanced judgments to maximize preservation of tendency information, and then predicts the annotator's label distributions for an input image ($\hat{p}_{A_i}$) supervised by multi-annotator dataset labels for training. This way, our learned annotator models better preserve the differences among annotators' tendencies, can gain more information about the annotator's tendency.

Compared to the existing methods, which typically build separate annotator models (shown in the left figure) for each annotator without considering inter-annotator correlations, which makes individual annotator modeling only rely on annotator labels for learning, our method \approach\ shared all specified annotators' queries (\ie, ${Q}_i$ in the right figure) in the Q-Former's self-attention module. This allows the model to capture inter-annotator correlations, providing an additional hidden supervisory signal from inter-annotator consistency to enhance modeling performance. Here, inter-annotator consistency reflects the agreement levels among annotators on given samples. Additionally, lightweight representing each annotator by query greatly reduces the cost of building separate conventional models for each annotator.

To evaluate how effectively the trained model preserves annotator tendency information, we provide a novel metric, the difference of inter-annotator consistency (DIC), which quantifies how inter-annotator correlations differ between ground-truths and predictions. To maximize the preservation of annotator tendency information, their correlation coefficients should be close, with smaller differences indicating that the model captures correlations and reflecting that the model well preserves the inherent patterns of disagreement among annotators.

Furthermore, we constructed two new large-scale datasets: STREET (city impression evaluation) and AMER (video emotion recognition), each providing substantial consecutive labels per annotator ID. In contrast, although crowdsourcing platforms have numerous annotators, only a small subset of samples are annotated consecutively with consistent annotator IDs \cite{SimLabel}. Our datasets offer valuable open-source resources to the community and further researchers. It is worth noting that AMER is the first multi-annotator multimodal dataset in this field. Our work makes the following contributions:

\begin{itemize}
  \item {\bf A new multi-annotator tendency learning task:} We introduce a new \task\ task that focuses on learning each annotator's tendency, better preserves tendency information. In contrast, existing tasks focus on consensus-oriented learning, learning a consensus label prediction via multi-annotator supervision, which easily averages out annotator differences and leads to tendency information loss.

  \item {\bf An efficient query-based baseline method for new task:} We propose a model \approach\ that uses lightweight queries to represent each annotator for tendency modeling and save costs of building separate conventional models for each annotator, leverages shared learnable queries to capture inter-annotator correlations as an additional hidden supervisory signal to enhance modeling performance.
  
  \item {\bf Two new large-scale datasets, a new metric:} We contribute two new datasets: STREET and AMER, providing averages of 4,300 and 3,118 per-annotator labels. We provide a new DIC metric to evaluate how effectively models preserve annotators' tendency information.
\end{itemize}

\begin{table*}[htbp]
    \centering
    \caption{Dataset comparison. Compared to existing multi-annotator datasets, our dataset contains a greater number of samples annotated by each annotator, which helps promote annotator tendency learning. Meanwhile, AMER is the first multimodal multi-annotator dataset.}
    \label{tab:dataset_comparison}
    \begin{tabular}{llcc}
    \toprule
    Dataset & Dataset description  & Modality & \# samples per annotator  \\
    \toprule
    QUBIQ-kidney \cite{menze2020quantification} & kidney image & image & 24 \\
    QUBIQ-tumor \cite{menze2020quantification} & brain tumor image & image & 32\\
    QUBIQ-growth \cite{menze2020quantification} & brain growth image & image & 39\\
    QUBIQ-prostate \cite{menze2020quantification} & prostate image  & image & 55\\
    CIFAR-10H \cite{peterson2019human} & object recognition  & image & 200 \\
    MUSIC \cite{rodrigues2013learning} & music genre classification  & audio & 2$\sim$368\\
    MURA \cite{rajpurkar2017mura} & radiographic image & image & 556 \\
    RIGA \cite{almazroa2017agreement} & retinal cup and disc segmentation & image & 750\\
    LIDC-IDRI \cite{armato2011lung} & lung nodule image  & image & 1,018\\
    \midrule
    \textbf{STREET (Ours)} & city impression evaluation & image  & 4,300 \\
    \textbf{AMER (Ours)} & video emotion recognition & audio, video, text  & 970$\sim$5,202\\
    \bottomrule
\end{tabular}
\end{table*}

\section{Related work}
\subsection{Multi-annotator Tendency Learning Task}
To the best of our knowledge, the multi-annotator tendency learning task has not yet been investigated. Traditional multi-annotator learning tasks focus on estimating consensus or ground-truth labels from multiple noisy annotations. These include early probabilistic models \cite{DS_model}, EM algorithms \cite{GLAD}, Gaussian models \cite{GP-MLL}, and biased estimation \cite{bias_annotator}. Tanno et al. \cite{tanno2019learning} proposed modeling annotator confusion matrices as learnable parameters in neural networks. Cao et al. \cite{cao2019max} introduced max-MIG to learn from multiple annotators. NEAL \cite{NEAL} employs neural expectation-maximization to jointly learn annotator expertise and true labels. Later methods used probabilistic frameworks to aggregate multiple annotations into a consensus or ground-truth label by confusion matrix \cite{Sampling-CM}, agreement distribution \cite{Learn2agree}, and Gaussian distributions \cite{PADL}. This consensus-oriented learning often easily averages out annotator differences and leads to tendency information loss. In contrast, our introduced \task\ task focuses on learning each annotator's tendency, better preserving tendency information.

\subsection{Annotator Tendency Learning Framework}
Previous studies typically construct independent models for each annotator to learn their specific behavioral patterns \cite{Confident}, or build separate distributions based on input features to fit each annotator \cite{PADL}, or use individual confusion matrices to approximate each annotator’s tendencies \cite{CNN-CM}. Recent research associates annotator influences with model outcomes. TAX \cite{TAX} explains pixel-level assignment decisions for semantic segmentation by associating learned convolutional kernels with a prototype library representing annotator tendencies. MAGI \cite{MAGI} leverages multiple annotators' explanations addressing challenges of noisy annotations and missing annotator correspondences. Schaekermann et al.~\cite{structured_adjudication} revealed how factors like expert background and data quality contribute to annotation disagreements. These approaches build separate conventional models for each annotator without considering inter-annotator correlations, which makes individual annotator modeling only rely on annotator labels for learning. In contrast, our method \approach\ uses lightweight queries to represent each annotator for tendency modeling and save the costs of building separate conventional models for each annotator. It also leverages shared learnable queries to capture inter-annotator correlations as an additional hidden supervisory signal to enhance modeling performance.

\subsection{Multi-annotator Datasets}
Most existing multi-annotator datasets often have only a small subset of samples with consecutive annotations from consistent annotator IDs. CIFAR-10H \cite{CIFAR-10H}, based on the CIFAR-10 \cite{CIFAR-10} dataset, includes 10,000 test samples labeled by 2,571 annotators, but each annotator ID has on average only about 200 consecutive labels. LabelMe \cite{Labelme} includes an average of approximately 42.4 consecutive labels per annotator ID. Audio dataset Music \cite{Music} contains an average of about 46.1 consecutive labels. The medical datasets are commonly used in multi-annotator studies, consecutive annotator labels are even sparser. QUBIQ \cite{QUBIQ}, a dataset for quantifying uncertainty in biomedical image segmentation, includes four distinct segmentation datasets with an average of only 40 samples, and even then, annotator IDs have only around 8 consecutive labeled samples each. More consecutive individual annotator labels facilitate \task, we contribute two large-scale datasets: STREET and AMER, averaging 4,300 and 3,118 consecutive labels per annotator ID, respectively. These datasets contribute significant open-source resources that we believe will benefit the research community.

\begin{figure*}[t]
  \centering
  \includegraphics[width=0.8\linewidth]{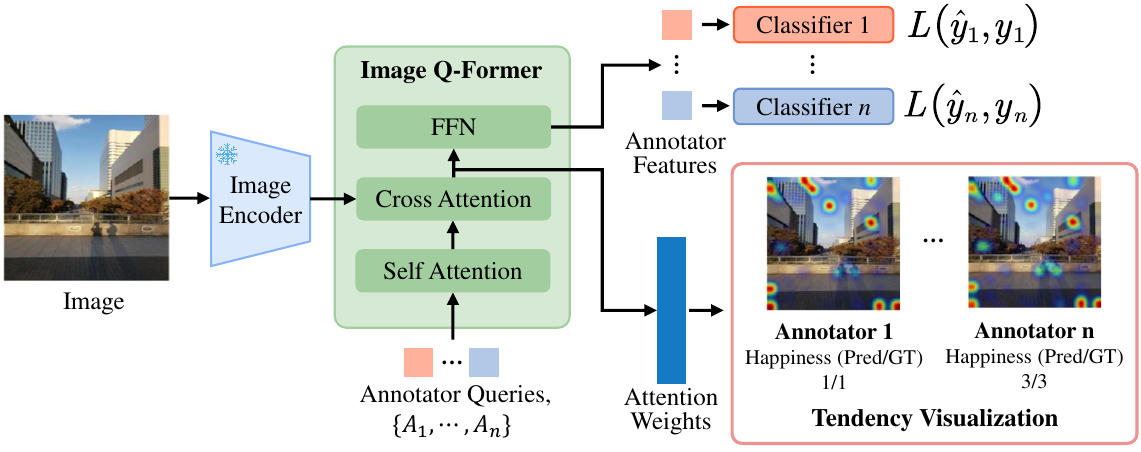}
    \caption{\approach\ architecture for image-specific pipeline. A frozen pre-trained image encoder extracts features, which interact with annotator-specific learnable queries in the Image Q-Former through cross-attention, producing annotator-specific features for classification. Different annotators’ cross-attention weights represent different annotator tendencies and provide visualization for analysis.}
    \label{fig:architecture-image}
\end{figure*}

\section{Dataset Construction}
\label{sec:dataset}
We contribute two new large-scale datasets: STREET (city impression assessment) and AMER (multi-modal emotion recognition) for our introduced \task\ task. Table \ref{tab:dataset_comparison} compares the current multi-annotator datasets. We observe that in existing datasets, the number of samples annotated by each annotator is relatively small, and there is a lack of multi-annotator multimodal datasets. For example, in the RIGA dataset \cite{almazroa2017agreement}, each annotator labels 750 samples, while in the CIFAR-10H dataset \cite{peterson2019human}, each annotator labels 200 samples.

{\textbf{(1) STREET}} is an urban perception dataset with multi-annotators, which contains $4,300$ high-resolution images covering various urban elements, such as streets, public spaces, and infrastructure. The images were captured during a series of city strolling surveys, which aim to analyze emotions in relation to various factors associated with the city. The surveys were conducted by an organization to which one of our co-authors belongs. Voluntary participants walked around their own familiar city, took photos of various factors that may affect their subjective feelings (i.e., happiness/health), and assigned labels related to these feelings to each image (though we do not use these labels but ones assigned by crowd workers in our experiments). Thirteen survey sessions were conducted in five different cities (three urban areas and two suburban areas). A total of 327 participants, ranging in age from their 10s to 60s, took part in the survey. Each session lasted about one hour, with each participant taking an average of 12.6 photos. We outsourced the annotation process to a company, which selected $10$ annotators with balanced age, gender, and location diversity on a platform similar to Amazon Mechanical Turk, assessing five perception dimensions: happiness, healthiness, safety, liveliness, and orderliness, using a $6$-point scale ($-3$ to $+3$). Each annotator spent approximately three weeks on their annotations. This multi-annotator dataset provides comprehensive human perception data for urban environments, enabling the quantitative analysis of environmental features and their emotional impact.

{\textbf{(2) AMER}} is a multimodal emotion dataset; the raw data is sourced from MER2024 \cite{lian2024mer}, which contains $5,207$ video samples from movies and TV series, with multi-annotator emotion labels \cite{MicroEmo-arxiv, MicroEmo-mm}. Each sample typically contains one person, with relatively complete speech content. We annotated AMER using the open-source software Label Studio \cite{LabelStudio}. We hired $15$ annotators, who were students of our co-author's institution, and underwent a training session with $10$ samples. We retained $13$ annotators after screening out careless and irresponsible ones. Each annotator completed the task in approximately two weeks, with scheduled breaks to maintain annotation quality, where each annotator selects the most likely label from 8 candidate labels, i.e., worry, happiness, neutrality, anger, surprise, sadness, other, and unknown. Among all annotators, $10$ annotators show consistent participation, each providing approximately $970$ to $1,096$ labels, while the remaining 3 annotators contribute over $5,000$ labels each. This rich multi-annotator setup provides reliable emotion annotation results and allows for a robust evaluation of emotion recognition performance.  

\section{Methodology}
\label{sec:method}
We propose \approach\ method for introduced new \task\ task. We take the image classification task to illustrate the image-specific architecture, which consists of a pre-trained image encoder, the annotator-specific learnable queries representing the tendencies of multi-annotators, an image Q-Former, and the classifiers for corresponding annotators, as shown in Figure~\ref{fig:architecture-image}. Note that this architecture is specifically designed for image inputs from the STREET dataset. For experiments on the AMER video dataset, we have a different video-specific architecture, which is detailed in the supplementary material.

As Fig.~\ref{fig:architecture-image} illustrates, given an image input $I \in \mathbb{R}^{H \times W \times 3}$, a frozen pre-trained image encoder \cite{EVA-CLIP} first extracts image features, which are then input to the Image Q-Former \cite{BLIP2}.
Subsequently, we define the unique tendency for annotator $A_k$ ($k = 1, \dots, n$) through each learnable query in a Q-Former, then all annotator-specific queries in cross-attention simultaneously interact with input features to capture their diverse tendencies and obtain annotator-specific feature. Finally, the specific annotator features are mapped through a fully connected layer to an appropriate feature dimension and then connected to each annotator’s corresponding classifier to output classification results. Note that although we assign a separate query for each annotator, they are processed by a multi-head attention module, typically with 12 heads. This setup provides each query with diverse perspectives. Our experiments have verified that the learned tendency information is as expected and effective.

This attention-based architecture design assumes that differences between annotators are attributed to what observers focus on, leverages learnable queries of Q-Former to model tendencies of different annotators, and uses a cross-attention module to interact with input features and capture different annotators' nuanced judgments to maximize preservation of tendency information. Meanwhile, all specified annotators' queries are shared in the self-attention module, allowing the model to capture inter-annotator correlations. This provides additional supervision signals from hidden inter-annotator consistency to enhance modeling performance. Here, inter-annotator consistency reflects the agreement levels among annotators on given samples. Additionally, lightweight representing each annotator by query greatly reduces the cost of building separate conventional models for each annotator.

For visualization, the cross-attention weights reflecting different focus points of annotators based on their tendencies can enable effective visualization for analysis. As shown in Figure~\ref{fig:architecture-image}, the focus points to image input are based on all patches of the image. Different annotators’ varying levels of focus on the specific semantics through different patches indicate their tendency differences: compared to \emph{annotator 1}, \emph{annotator n} exhibits a higher level of focus on the silhouettes of the two people holding hands in the image. This dataset is for city impression emotion classification. Analyzing the annotation and results reveals interesting patterns: \emph{annotator n} labeled a higher score than Annotator 1 to the ``happiness'' dimension, a difference that could potentially be linked to their varying levels of focus on specific image patches, which might have influenced their judgments to some extent. This further suggests that \task\ is an intriguing task with notable potential research value.

\subsection{Loss Function}
Finally, as shown in Figure~\ref{fig:architecture-image}, the total training loss $\mathcal{L}_{\text{total}}$ for the proposed multi-annotator classification model, \approach, is defined as the sum of individual cross-entropy losses for each annotator:
\begin{equation}
\mathcal{L}_{\text{total}} = \sum_{k=1}^{n} \mathcal{L}(\hat{y}_{k}, y_{k}),
\end{equation}
where each annotator $A_{k}$ has a specific predicted probabilities $\hat{y}_{k} \in [0, 1]^C$, a reference label $y_{k} \in \{0, 1\}^C$ in one-hot vector representation, and $C$ is the number of classes.

\section{Experiment}
\label{sec:experiment}
We conduct extensive experiments to compare our \approach\ with competing approaches, focusing on individual annotator modeling, inter-annotator consistency, assessing their utility for consensus prediction, testing applicability under sparse annotations, and complemented with visualization analysis. To verify the effectiveness of \approach, we select three representative baselines: D-LEMA \cite{D-LEMA} is an ensemble architecture for multi-annotator learning; PADL \cite{PADL} uses Gaussian distribution fitting each annotator's tendency; MaDL \cite{MaDL}, which employs individual confusion matrices to approximate each annotator's tendency. Meanwhile, to verify that the model has truly learned the tendency information rather than the features from the encoder, we specifically set a Base baseline that only retains the encoder and classifiers as a reference. Except for Accuracy and $F_1$ score \cite{F1-score} metrics, we evaluate and discuss results on our two novel datasets using the new DIC metric. Note that additional experiments such as model efficiency and additional results, along with extended discussions, are provided in the supplementary material due to space limitation.

\begin{table*}
\centering
\caption{The accuracy metric is to evaluate results on the STREET dataset. We assess performance for individual annotator modeling (each annotator $A_{k}$, \textit{k} = 1, \dots, 13), the average (Avg), and consensus prediction (CoPr). Higher is better.}
\label{tab:street_accuracy}
\begin{tabular}{llcccccccccccc}
\toprule
Perspectives & Methods & $A_1$ & $A_2$ & $A_3$ & $A_4$ & $A_5$ & $A_6$ & $A_7$ & $A_8$ & $A_9$ & $A_{10}$ & Avg & CoPr \\
\midrule
\multirow{4}{*}{Happiness} & Base & 0.80 & 0.12 & 0.27 & 0.38 & 0.56 & 0.35 & 0.44 & 0.44 & 0.31 & 0.55 & 0.42 & 0.45 \\
& D-LEMA & 0.85 & 0.71 & 0.44 & 0.36 & \textbf{0.70} & 0.43 & 0.46 & 0.54 & 0.41 & 0.47 & 0.54 & 0.57 \\
 & PADL & 0.93 & 0.74 & 0.48 & 0.53 & 0.57 & 0.47 & 0.42 & 0.51 & 0.50 & 0.60 & 0.58 & 0.55 \\
 & MaDL & 0.91 & 0.77 & 0.44 & 0.38 & \textbf{0.70} & 0.47 & 0.46 & \textbf{0.54} & 0.48 & 0.47 & 0.56 & 0.58 \\
 & Ours & \textbf{0.94} & \textbf{0.80} & \textbf{0.54} & \textbf{0.55} & 0.69 & \textbf{0.51} & \textbf{0.53} & \textbf{0.54} & \textbf{0.52} & \textbf{0.64} & \textbf{0.63} & \textbf{0.62} \\
\midrule
\multirow{4}{*}{Healthiness} & Base & 0.77 & 0.19 & 0.14 & 0.47 & 0.87 & 0.36 & 0.43 & 0.44 & 0.51 & 0.54 & 0.47 & 0.56 \\
 & D-LEMA & 0.83 & \textbf{0.77} & 0.44 & 0.43 & 0.87 & 0.41 & 0.44 & 0.46 & 0.55 & 0.46 & 0.57 & 0.54 \\
 & PADL & \textbf{0.92} & 0.72 & 0.55 & 0.44 & 0.84 & 0.47 & 0.46 & 0.44 & 0.52 & 0.55 & 0.59 & 0.50 \\
 & MaDL & 0.89 & \textbf{0.77} & 0.44 & 0.44 & \textbf{0.90} & 0.41 & 0.44 & 0.46 & 0.55 & 0.46 & 0.58 & 0.55 \\
 & Ours & \textbf{0.92} & 0.75 & \textbf{0.56} & \textbf{0.52} & \textbf{0.90} & \textbf{0.49} & \textbf{0.48} & \textbf{0.54} & \textbf{0.64} & \textbf{0.61} & \textbf{0.64} & \textbf{0.58} \\
\midrule
\multirow{4}{*}{Safety} & Base & 0.58 & 0.65 & 0.36 & 0.36 & 0.61 & 0.37 & 0.56 & 0.40 & 0.32 & 0.53 & 0.47 & 0.51 \\
 & D-LEMA & 0.62 & 0.69 & 0.27 & 0.41 & 0.50 & 0.46 & 0.48 & 0.40 & 0.35 & 0.50 & 0.47 & 0.49 \\
 & PADL & \textbf{0.72} & 0.78 & 0.24 & 0.44 & 0.69 & 0.44 & \textbf{0.53} & 0.42 & 0.46 & 0.48 & 0.52 & 0.54 \\
 & MaDL & 0.63 & 0.63 & 0.27 & 0.32 & 0.61 & 0.38 & 0.46 & 0.42 & 0.36 & 0.52 & 0.46 & 0.56 \\
 & Ours & \textbf{0.72} & \textbf{0.80} & \textbf{0.38} & \textbf{0.48} & \textbf{0.71} & \textbf{0.54} & \textbf{0.53} & \textbf{0.50} & \textbf{0.52} & \textbf{0.58} & \textbf{0.58} & \textbf{0.61} \\
\midrule
\multirow{4}{*}{Liveliness} & Base & 0.79 & 0.60 & 0.53 & 0.46 & 0.76 & 0.30 & 0.35 & 0.36 & 0.53 & 0.57 & 0.53 & 0.55 \\
 & D-LEMA & 0.79 & 0.58 & 0.37 & 0.42 & 0.74 & 0.38 & 0.44 & 0.41 & 0.50 & 0.46 & 0.51 & 0.53 \\
 & PADL & 0.85 & 0.66 & 0.56 & 0.46 & 0.75 & 0.44 & 0.43 & 0.47 & 0.56 & 0.57 & 0.58 & 0.54 \\
 & MaDL & 0.78 & 0.56 & 0.35 & 0.40 & 0.76 & 0.34 & \textbf{0.48} & 0.42 & 0.47 & 0.47 & 0.50 & 0.56 \\
 & Ours & \textbf{0.87} & \textbf{0.68} & \textbf{0.57} & \textbf{0.53} & \textbf{0.80} & \textbf{0.49} & \textbf{0.48} & \textbf{0.51} & \textbf{0.62} & \textbf{0.61} & \textbf{0.62} & \textbf{0.59} \\
\midrule
\multirow{4}{*}{Orderliness} & Base & 0.50 & 0.64 & 0.39 & 0.45 & 0.86 & 0.31 & 0.39 & 0.34 & 0.31 & 0.49 & 0.47 & 0.52 \\
 & D-LEMA & 0.55 & 0.60 & 0.32 & 0.36 & 0.82 & 0.39 & 0.42 & 0.36 & 0.37 & 0.47 & 0.47 & 0.57 \\
 & PADL & 0.73 & 0.65 & 0.44 & 0.45 & 0.93 & 0.45 & 0.45 & 0.36 & 0.42 & \textbf{0.63} & 0.55 & 0.54 \\
 & MaDL & 0.61 & 0.60 & 0.34 & 0.36 & 0.86 & 0.37 & 0.47 & 0.36 & 0.37 & 0.49 & 0.48 & 0.58 \\
 & Ours & \textbf{0.74} & \textbf{0.71} & \textbf{0.52} & \textbf{0.55} & \textbf{0.94} & \textbf{0.47} & \textbf{0.54} & \textbf{0.44} & \textbf{0.56} & 0.62 & \textbf{0.61} & \textbf{0.62} \\
\bottomrule
\end{tabular}
\end{table*}

\begin{table*}
\centering
\caption{The accuracy (ACC) and $F_1$ score evaluate results on the AMER dataset. We assess performance for individual annotator modeling (each annotator $A_{k}$, \textit{k} = 1, \dots, 13), the average (Avg), and consensus prediction (CoPr). Higher is better.}
\label{tab:amer_accuracy_f1}
\begin{tabular}{clccccccccccccccc}
\toprule
Metric & Methods & $A_1$ & $A_2$ & $A_3$ & $A_4$ & $A_5$ & $A_6$ & $A_7$ & $A_8$ & $A_9$ & $A_{10}$ & $A_{11}$ & $A_{12}$ & $A_{13}$ & Avg & CoPr \\
\midrule
\multirow{5}{*}{ACC} 
& Base & 0.30 & 0.29 & 0.43 & 0.25 & 0.24 & 0.20 & 0.26 & 0.19 & 0.34 & 0.10 & 0.41 & 0.48 & 0.19 & 0.28 & 0.35 \\
& D-LEMA & 0.86 & 0.88 & 0.85 & 0.87 & 0.89 & 0.86 & 0.88 & 0.87 & 0.85 & 0.86 & 0.45 & 0.51 & 0.33 & 0.78 & 0.55 \\
& PADL & 0.89 & 0.90 & 0.88 & 0.93 & 0.87 & 0.91 & 0.86 & 0.94 & 0.89 & 0.88 & 0.47 & 0.54 & 0.35 & 0.79 & 0.52 \\
& MaDL & 0.93 & 0.91 & 0.90 & 0.89 & 0.90 & 0.88 & 0.90 & 0.89 & 0.87 & 0.92 & 0.50 & 0.53 & 0.37 & 0.80 & 0.57 \\
& Ours & \textbf{0.94} & \textbf{0.93} & \textbf{0.93} & \textbf{0.94} & \textbf{0.94} & \textbf{0.92} & \textbf{0.93} & \textbf{0.95} & \textbf{0.93} & \textbf{0.93} & \textbf{0.59} & \textbf{0.61} & \textbf{0.40} & \textbf{0.84} & \textbf{0.60} \\
\midrule
\multirow{5}{*}{$F_1$} 
& Base & 0.26 & 0.27 & 0.40 & 0.22 & 0.23 & 0.17 & 0.24 & 0.18 & 0.31 & 0.08 & 0.35 & 0.41 & 0.14 & 0.25 & 0.32 \\
& D-LEMA & 0.84 & 0.87 & 0.81 & 0.84 & 0.86 & 0.85 & 0.86 & 0.82 & 0.83 & 0.84 & 0.38 & 0.44 & 0.27 & 0.73 & 0.52 \\
& PADL & 0.86 & 0.88 & 0.85 & 0.91 & 0.83 & 0.89 & 0.82 & 0.92 & 0.85 & 0.86 & 0.41 & 0.50 & 0.29 & 0.76 & 0.49 \\
& MaDL & 0.90 & 0.85 & 0.87 & 0.86 & 0.87 & 0.85 & 0.87 & 0.86 & 0.84 & 0.92 & 0.45 & 0.48 & 0.33 & 0.77 & 0.54 \\
& Ours & \textbf{0.91} & \textbf{0.91} & \textbf{0.90} & \textbf{0.92} & \textbf{0.91} & \textbf{0.90} & \textbf{0.89} & \textbf{0.93} & \textbf{0.91} & \textbf{0.93} & \textbf{0.54} & \textbf{0.55} & \textbf{0.34} & \textbf{0.81} & \textbf{0.57} \\
\bottomrule
\end{tabular}
\end{table*}

\subsection{Implementation Details}
For the model pipeline as shown in Figure~\ref{fig:architecture-image}, we use ViT-G/14 from EVA-CLIP \cite{EVA-CLIP} as the encoder, with the image Q-Former initialized from InstructBLIP \cite{InstructBLIP} (the Frame Q-Former is same, Video Q-Former is initialized from Video-LLaMA \cite{Video-llama} in video-specific pipeline from supplementary material). The input image and video are resized to 224$\times$224 and further normalized. The number of query tokens matches the number of annotators, and each annotator's classifier model uses an MLP. During training, we use the AdamW optimizer with an initial learning rate of 1e-4, weight decay of 0.01, and a maximum gradient norm of 1.0 for gradient clipping. A linear warmup strategy is applied for the first 20\% steps followed by cosine learning rate decay. We set the maximum number of epochs to 200, with early stopping (patience being 25) to prevent overfitting. To accelerate training, the model is trained using distributed data parallel (DDP) on four NVIDIA V100 GPUs.

\begin{figure*}
  \centering
  \includegraphics[width=0.85\linewidth]{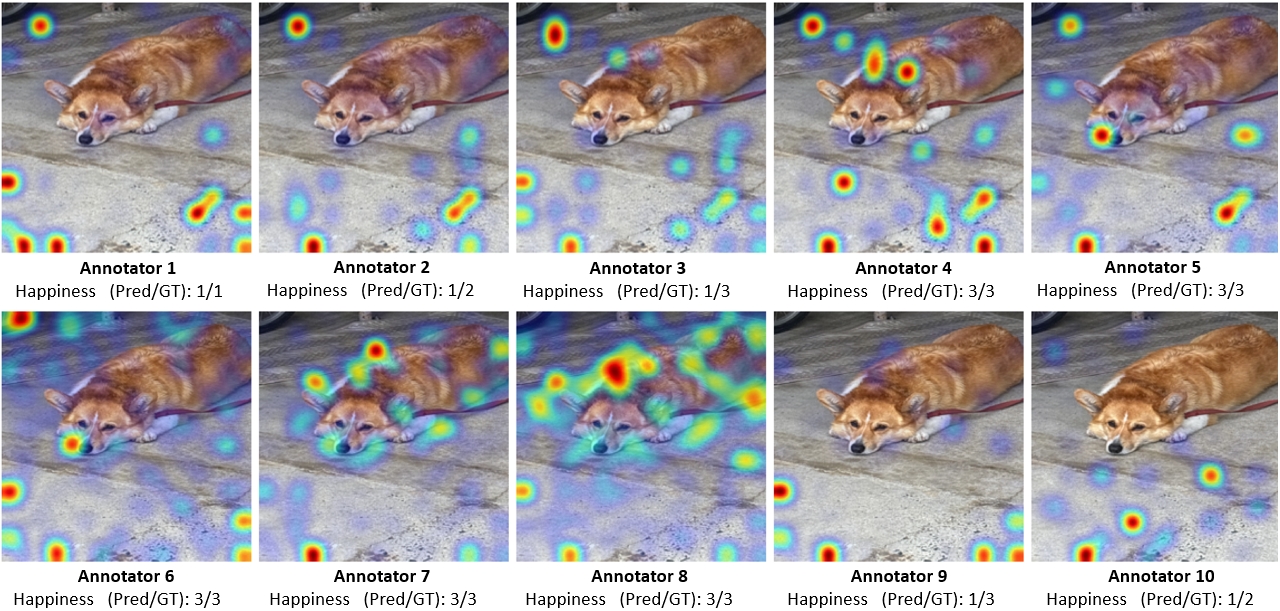}
  \caption{We provide the visualization analysis of the annotator tendencies on the STREET dataset. The focuses of 10 annotators reveal distinct preferences, where \emph{annotators 4, 5, 6, 7, and 8} exhibit centralized focuses on a cute dog compared to other annotators.}
  \label{fig:ii}
\end{figure*}

\begin{figure*}
  \centering
  \includegraphics[width=0.85\linewidth]{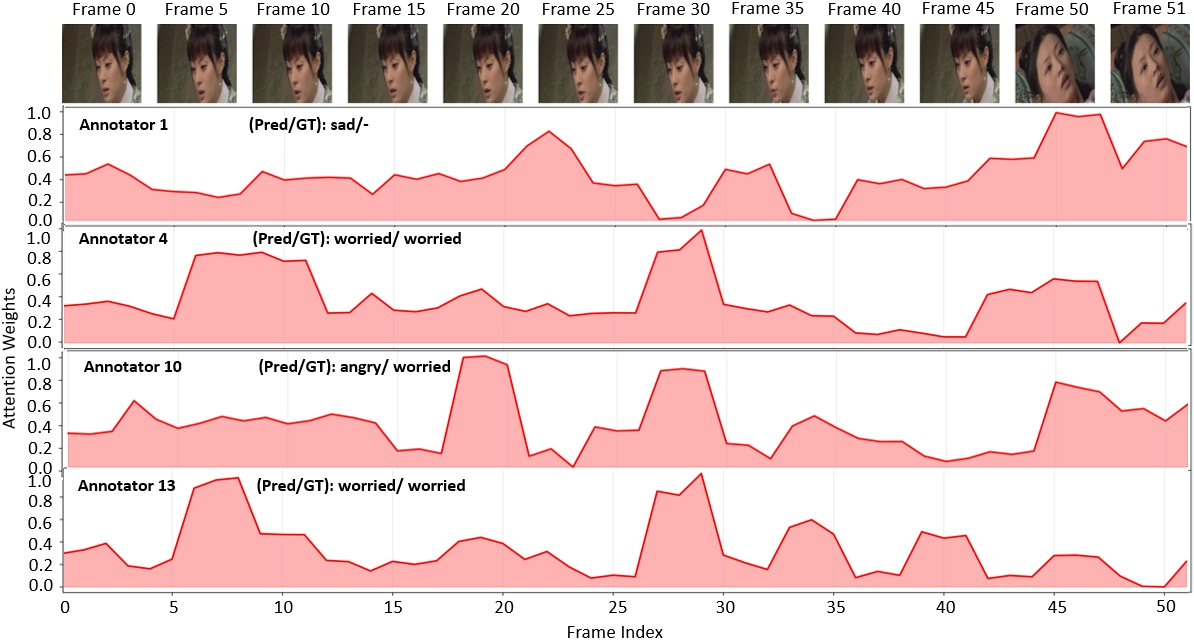}
  \caption{We provide the visualization analysis of the annotator tendencies on the AMER dataset. The \emph{annotator 1} exhibits focus on final frames (the frames contain different person), while \emph{annotators 4 and 13} focus on the middle frames.}
  \label{fig:vv}
\end{figure*}

\subsection{Evaluation Metrics}
For evaluating \task\ task, accuracy is the fundamental metric for the performance of individual annotator modeling and consensus prediction (majority-votes the predictions of multi-annotators). We also use $F_1$ score \cite{F1-score} balancing precision and recall, suitable for potential class imbalance from uneven annotation densities, as in AMER (1,040 vs. 5,195 labels for annotators 1–10 vs. 11–13). To evaluate how effectively trained models preserve annotators' tendency information, we propose a new metric, coined difference of inter-annotator consistency (DIC).

\textbf{Difference of inter-annotator consistency (DIC)} quantifies how inter-annotator correlations differ between ground-truths and predictions. To maximize the preservation of annotator tendency information, their correlation coefficients should be close, with smaller differences indicating that the model captures correlations and reflects that the model well preserves the inherent patterns of disagreement among annotators. The annotations given by different annotators differ, and their consistency can be measured by Cohen's kappa coefficient \cite{kappa}. Let $Y_k$ and $Y_{l}$ denote the set of all annotations by annotators $A_k$ and $A_{l}$. We can compute \textit{inter-annotator consistency matrix} $M$ whose $(k,l) $-th element $m_{kl}$ is given by
\begin{align}
    m_{kl} = \kappa(Y_k, Y_l),
\end{align}
where $\kappa(Y_k, Y_l)$ gives Cohen's kappa coefficient for $Y_k$ and $Y_l$. We can also compute the inter-annotator consistency matrix $M'$ but for predictions $\hat{Y}_k$ and $\hat{Y}_l$, where its $(k,l)$-th element $m'_{kl}$ is given by
\begin{align}
    m'_{kl} = \kappa(\hat{Y}_k,\hat{Y}_l). 
\end{align}
The better the model preserves the tendency information and fits patterns, the smaller the difference between these matrices. We quantify it by
\begin{align}
    \text{DIC} = \|M-M'\|_F,
\end{align}
where $\|\cdot\|_F$ denotes the Frobenius norm.
A lower DIC indicates that the model better preserves the annotator tendency information and successfully captures each annotator's tendency (with respect to the others).

\begin{table}
\centering
\caption{DIC scores to evaluate inter-annotator tendency discrepancy patterns on STREET and AMER datasets, lower values indicate better tendency preservation. -Ha, -He, -Sa, -Li, and -Or represents five perspectives of STREET dataset: happiness, healthiness, safety, liveliness, and orderliness.}
\label{tab:consistency}
\begin{tabular}{lcccc}
\toprule
Datasets & D-LEMA & PADL & MaDL & Ours \\
\midrule
S-Ha & 0.62 & 0.48 & 0.45 & \textbf{0.43} \\
S-He & 0.59 & 0.52 & 0.45 & \textbf{0.38} \\
S-Sa & 0.36 & 0.32 & 0.29 & \textbf{0.24} \\
S-Li & 0.51 & 0.43 & 0.39 & \textbf{0.27} \\
S-Or & 0.61 & 0.57 & 0.59 & \textbf{0.54} \\
AMER & 0.42 & 0.36 &  0.31 & \textbf{0.23} \\
\bottomrule
\end{tabular}
\end{table}

\subsection{Quantitative Results}
For individual annotator modeling, results in Tables~\ref{tab:amer_accuracy_f1} and \ref{tab:street_accuracy} show that our method consistently outperforms all baselines in both accuracy and $F_1$ score (See supplementary material for $F_1$ results on STREET dataset) across individual annotators on the STREET and AMER datasets.
This validates the superiority of our approach in capturing individual annotator behavior patterns.

For inter-annotator consistency, the DIC scores in Table~\ref{tab:consistency} also show that our method surpasses baseline approaches on both STREET and  AMER datasets.
We indirectly quantify how well the model preserves the annotators' tendency information through the inter-annotator consistency matrices.
These comparative results verify the superiority of our proposed method in modeling individual annotators and the ability to better preserve annotators' tendency information.

For consensus application benefits, real-world applications often seek a single consensus label despite subjectivity among annotators. We evaluate consensus prediction to validate whether modeling individual annotators preserves valuable information for practical needs. As no definitive ground truth exists, we use majority vote over raw annotations as a proxy, acknowledging its potential biases.
Existing baselines adopt different aggregation strategies: D-LEMA learns weighted fusion; PADL applies meta-learning; and MaDL jointly optimizes consensus and annotator classifiers. They may average annotator perspectives during training, potentially diminishing individual nuances.
For a fair comparison, we apply unified majority voting over annotator-specific predictions from all methods rather than using their original aggregated outputs. Results (CoPr) in Tables~\ref{tab:amer_accuracy_f1}, \ref{tab:street_accuracy}, and $F_1$ results show that our method achieves superior consensus performance, suggesting that modeling individual annotators helps retain valuable information, potentially benefiting real-world consensus applications.
\textit{Note:} This experiment serves to validate practical utility of individual annotator modeling rather than to assert overall superiority.

For applicability under sparse annotations, to evaluate our model’s applicability under sparse annotation scenarios, we simulated real-world conditions by randomly removing annotations at various rates. As shown in Table~\ref{tab:sparse}, when 40\% of annotations are removed (See supplementary material for results of more sparse rates), our model’s average performance drops by 20.4\%, whereas the best baseline PADL experiences a larger drop of 27.4\%.
Results suggest that our superiority stems from modeling inter-annotator correlations, which regularizes individual annotator representations, preventing overfitting to sparse labels and promoting consistency with shared patterns across annotators, to enhance robustness and generalization under sparse annotations.

\subsection{Qualitative Results}
We qualitatively analyze and discuss the visualization of annotator tendencies, which are exhibited by cross-attention weights of Q-former reflecting different focus points of annotators based on their tendencies.

As shown in Figure \ref{fig:ii}, on the STREET dataset, different annotators’ varying levels of focus on the specific semantics through different patches indicate their tendency differences: \emph{annotator 4, 5, 6, 7, and 8} centralized focus more on a cute dog in the image, while other annotators show less focus on this element. This dataset is for city impression emotion classification. Analyzing the annotation and visualization results reveals: \emph{annotator 4, 5, 6, 7, and 8} labeled higher scores to the ``happiness'' dimension, a difference which could potentially be linked to their varying levels of focus on specific semantic of image patches, which might have influenced their judgments. As shown in Figure \ref{fig:vv}, on the AMER dataset, annotators demonstrate diverse preferences through their varying focus on different frames: \emph{annotator 1} shows higher focus on the final frames (45-50), while \emph{annotator 4 and 13} focuses more on the early frames (5-10), \emph{annotator 10} shows a different focus. This dataset is video emotion classification. Analyzing the annotation and visualization results reveals that \emph{annotator 1} predicted ``sad'', while \emph{annotator 4 and 13} predicted ``worried''. The differences in the person or conversations of frame segments they focused on might partially contribute to the variations in their judgments.

\begin{table}[htbp]
\centering
\caption{Evaluation by accuracy for sparse scenarios (40\% of annotations are randomly removed). S-Ha, S-He, S-Sa, S-Li, and S-Or represent five perspectives of STREET dataset: happiness, healthiness, safety, liveliness, and orderliness.}
\label{tab:sparse}
\begin{tabular}{l@{\hspace{2.8mm}}ccccc@{\hspace{3mm}}c}
\toprule
Method & S-Ha & S-He & S-Sa & S-Li & S-Or & AMER \\
\midrule
Full-PADL & 0.58 & 0.59 & 0.52 & 0.58 & 0.55 & 0.79 \\
Full-Ours & 0.63 & 0.64 & 0.58 & 0.62 & 0.61 & 0.84 \\
\midrule
Sparse-PADL & 0.43 & 0.42 & 0.38 & 0.41 & 0.40 & 0.58 \\
Sparse-Ours & \textbf{0.52} & \textbf{0.51} & \textbf{0.46} & \textbf{0.49} & \textbf{0.48} & \textbf{0.66} \\
\bottomrule
\end{tabular}
\end{table}

\begin{table}[htbp]
\centering
\caption{Ablation study. The average performance of replacing some modules for annotator modeling, removing inter-annotator correlations, as well as a comparison of label aggregation with and without annotator tendency modeling, evaluated by accuracy.}
\label{tab:ablation}
\begin{tabular}{llccccccccccc}
\toprule
Method & S-Ha & S-He & S-Sa & S-Li & S-Or & AMER \\
\midrule
Base & 0.42 & 0.47 & 0.47 & 0.53 & 0.47 & 0.28 \\
w/ U-cls & 0.50 & 0.56 & 0.49 & 0.56 & 0.49 & 0.61 \\
w/o S-Attn & 0.52 & 0.54 & 0.46 & 0.53 & 0.56 & 0.73 \\
\midrule
Pre-mv & 0.58 & 0.57 & 0.56 & 0.58 & 0.60 & 0.43 \\
Post-mv & 0.62 & 0.58 & 0.61 & 0.59 & 0.62 & 0.60 \\
\midrule
Full (avg) & 0.63 & 0.64 & 0.58 & 0.62 & 0.61 & 0.84 \\
\bottomrule
\end{tabular}
\end{table}

\subsection{Ablation Study}
\label{sec:ablation_study}
We conduct an ablation study for model structure verification. First, we remove Q-Former reducing to the Base model; Second, we only use a unified classifier (w/ U-cls). As shown in Table~\ref{tab:ablation}, either of these decreases the overall annotator modeling performance, highlighting their effectiveness in our architecture. To verify the effectiveness of inter-annotator correlations as additional supervision enhancing modeling performance, we remove the self-attention module (w/o S-Attn) in Q-Former. Experimental results show that this decreases the performance, which demonstrates the advantage of capturing inter-annotator correlations in our architecture design. In addition, we also compared the majority voting consensus label training results (Pre-mv) with the majority voting results after modeling individual annotators (Post-mv). For Pre-mv, we perform majority voting on multiple annotator labels to obtain a label $L_{mv}$ and then apply global average pooling on whole queries of Q-Former to a single classifier for end-to-end training to obtain a single inference result $R_{pre}$. For Post-mv, we model annotators' tendency by learnable queries to independent classifiers for training and then integrate all annotators' inference results through majority voting to obtain a single result $R_{post}$. Evaluation results comparing $R_{pre}$ and $R_{post}$ with $L_{mv}$ show that Post-mv outperforms Pre-mv overall, validating our insight that modeling tendencies of individual annotators may lead to better specific task results.

\section{Conclusion}
This paper introduces a new Multi-annotator Tendency Learning (\task) task, proposes a baseline method \approach\ for this task, provides a new DIC metric, and constructs two large-scale datasets: AMER and STREET. Compared to existing multi-annotator tasks causing the tendency information loss, \task\ better preserves the tendency information. Our proposed method captures inter-annotator correlations, which provide an additional supervisory signal from hidden inter-annotator consistency to enhance annotator modeling performance compared to other models. Extensive experiments verified the effectiveness of our method.

\begin{figure*}[t]
  \centering
  \includegraphics[width=0.8\linewidth]{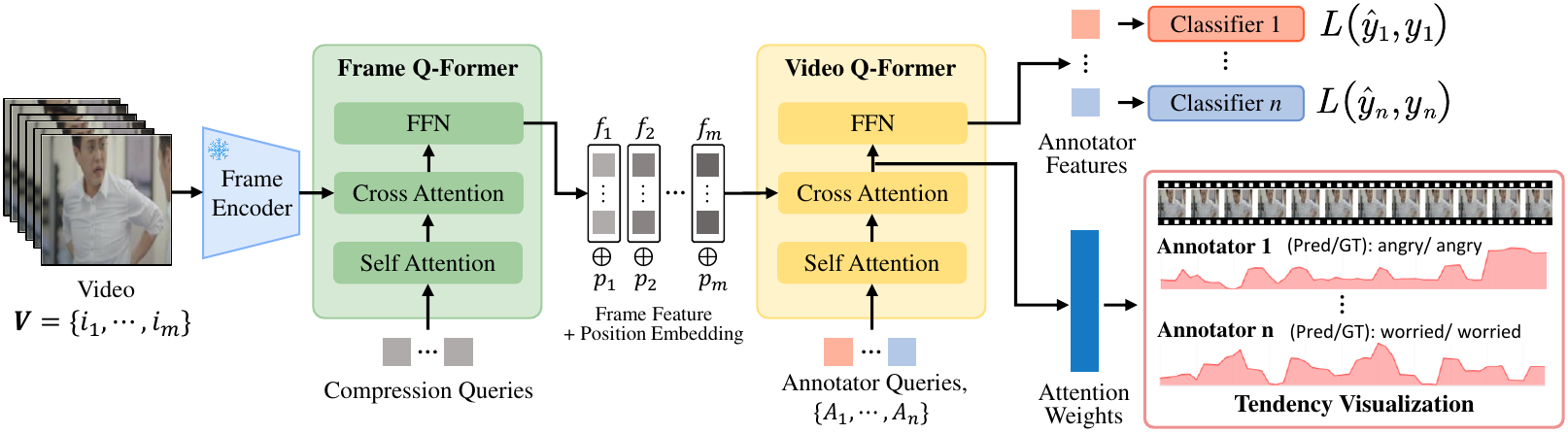}
    \caption{\approach\ architecture for video-specific pipeline. A frozen pre-trained encoder extracts frame features, which are compressed through the Frame Q-Former. With added frame position embedding, these interact with $1, \dots, n$ annotator-specific learnable queries in the Video Q-Former through cross-attention, producing annotator-specific features for classification. Different annotators’ cross-attention weights represent different annotator tendencies and provide visualization for analysis.}
    \label{fig:architecture-video}
\end{figure*}

\section{Supplementary Material}
\subsection{Overview}
In this supplementary, we first illustrate the detailed video-specific architecture in Section \ref{sec:architecture}.
We provide the analysis of model efficiency in Section \ref{sec:efficiency}.
Also, the extended discussion is provided in Section \ref{sec:discussion}.
Finally, we show the additional experimental results in Section \ref{sec:results}.

\subsection{Video-specific Architecture}
\label{sec:architecture}
Different from the image input pipeline in the main paper, Figure~\ref{fig:architecture-video} presents the video input pipeline specifically designed for the  AMER, etc. video datasets. Specifically, given a video input $V \in \mathbb{R}^{T \times H \times W \times 3}$, a frozen pre-trained image encoder \cite{EVA-CLIP} first extracts frame features, which are then further compressed by the image Q-Former using a specific number of compression queries, typically 32, to alleviate subsequent computational costs \cite{BLIP2}. The frame position embedding is then added and input to the video Q-Former. Subsequently, we define the unique tendency for annotator $A_k$ ($k = 1, \dots, n$) through each learnable query in a Q-Former, then all annotator-specific queries in cross-attention simultaneously interact with input features to capture their diverse tendencies and obtain annotator-specific feature. Finally, the specific annotator features are mapped through a fully connected layer to an appropriate feature dimension and then connected to each annotator’s corresponding classifier to output classification results.

For visualization, the focus points to video input are based on all frames of the video. Different annotators’ varying levels of focus on different frames indicate their tendency differences: \emph{annotator 1} shows more focus on the final frames, while \emph{annotator n} focuses more on the middle frames. This dataset is video emotion classification. Analyzing the annotation and visualization results, we see that annotators labeled emotions ``anger'' and ``worried''. The differences in the frame segments they focused on might partially contribute to the variations in their judgments.

\begin{table}[htbp]
\centering
\small
\caption{Model efficiency analysis. In this table, we report the number of parameters and the average inference time per sample. Lower values for both metrics indicate higher model efficiency.}
\label{efficiency}
\begin{tabular}{cccccc}
\toprule
Models & Parameters\thinspace(M) & Average Processing Time\thinspace(s) \\
\midrule
D-LEMA & $214.18$ & $5.64$ \\
PADL & $168.94$ & $4.59$ \\
MaDL & $201.37$ & $5.06$ \\
Ours & $\bf{106.02}$ & $\bf{4.28}$ \\
\bottomrule
\end{tabular}
\end{table}

\subsection{Model Efficiency Analysis}
\label{sec:efficiency}

This section evaluates the model efficiency from two aspects: model complexity and inference time. For model complexity, we use the number of parameters as the evaluation metric; for inference time, we compute the average processing time per sample \cite{Panoptic-tcsvt, Panoptic-wacv, Thermal-to-Color, phd}. To ensure a fair comparison, Table \ref{efficiency} compares the performance of different methods in the image input pipeline (see Figure \ref{fig:architecture-image} in the main paper). Specifically, we evaluate the efficiency of different methods on one perspective of the  STREET dataset with 10 annotators. It is worth noting that different methods use different backbone networks. To ensure fairness, we standardize the backbone network to ResNet-34 for all methods. In Table \ref{efficiency}, we observe that our model has fewer parameters than the other models, while also maintaining competitive average processing time. This validates our claim in the paper that, compared to the baselines that create separate conventional models for each annotator, our architecture demonstrates significant advantages in model efficiency.


\subsection{Discussion}
\label{sec:discussion}
We supplement some extended discussions about our work here.
Our architecture design superiorly balances effectiveness and complexity, provides an accompanying visualization analysis of annotator tendencies through cross-attention weights. This has potential value and interest for understanding different annotator judgments as described in the main paper. Currently, it is introduced as an accompanying function rather than a main contribution of our paper, but in the future, we hope it can be developed into a mature interpretability solution. We plan to explore to enhance it, which may need pixel-level annotations indicating specific regions that annotators focus on during the annotation process, although this could be expensive. We could also explore quantitative evaluation ways to this further interpretability, such as feature importance ranking consistency, attention-based fidelity metrics, and human intuition consistency assessments through user studies, etc, to further enhance its value and contribution to the community.

In the ablation study of our main paper, we validated an insight that modeling multiple annotator tendencies can obtain more annotator information and demonstrated its effectiveness through experimental results based on consensus aggregation using majority voting. In future work, we plan to extend more similar experimental scenarios and validations to enhance the explanation of how multi-annotator tendency learning helps and adds benefits to specific applications \cite{uneven}.

\subsection{Additional Results}
\label{sec:results}
We provide additional experimental results about visualization analysis of annotator tendencies on both AMER and STREET datasets.

\begin{figure*}
  \centering
  \begin{subfigure}{1.0\linewidth}
    \centering
    \includegraphics[width=0.85\linewidth]{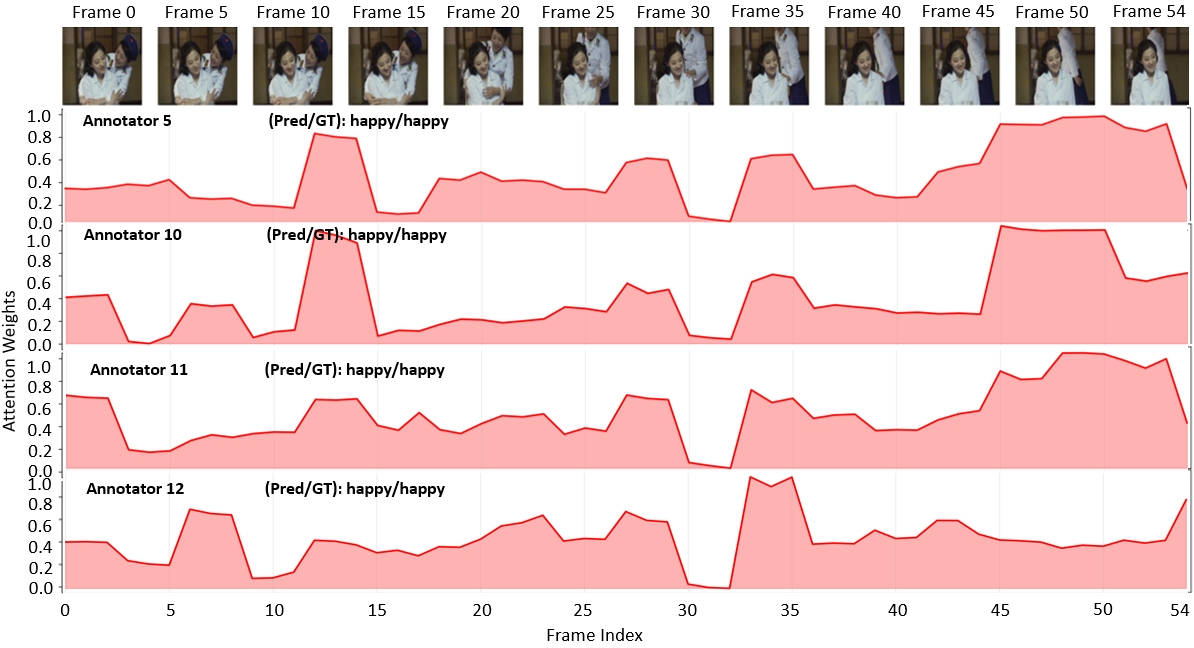}
    \caption{Sample 1.}
    \label{fig:va1}
  \end{subfigure}

  \vspace{2em}

  \begin{subfigure}{1.0\linewidth}
    \centering
    \includegraphics[width=0.85\linewidth]{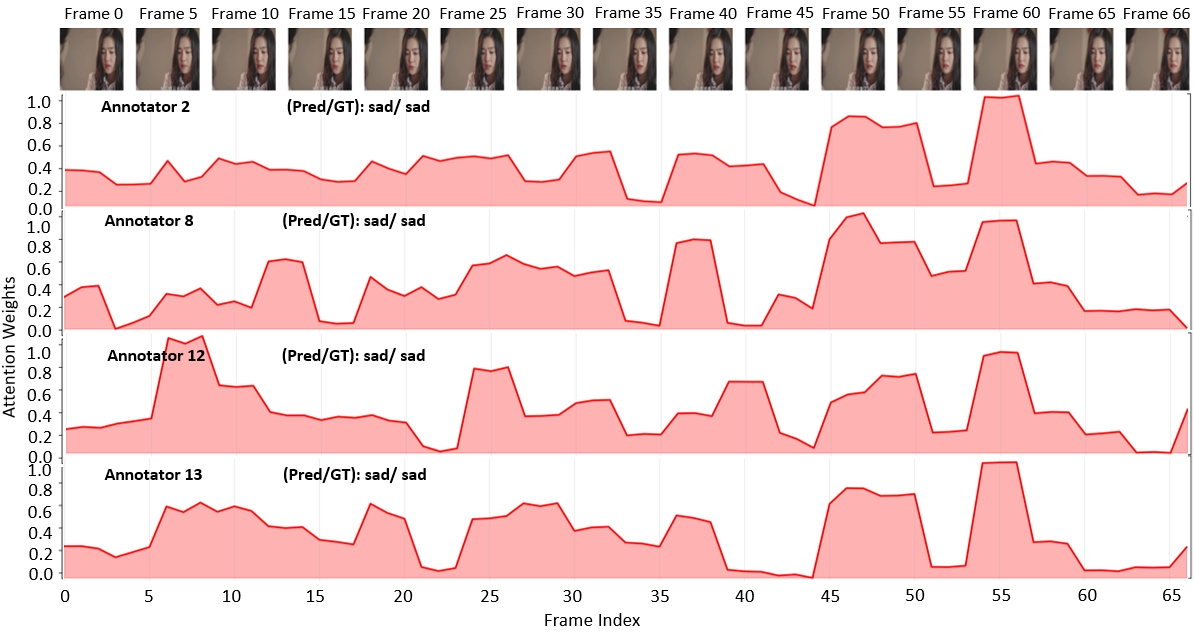}
    \caption{Sample 2.}
    \label{fig:va2}
  \end{subfigure}

  \caption{Additional sample experimental results visualize the annotator tendencies on the AMER video dataset. The tendencies of multi-annotators reveal distinct preferences.}
  \label{fig:va}
\end{figure*}

As shown in Figure~\ref{fig:va}, on the AMER dataset, annotators demonstrate different tendencies through varying focus on different frames. Sample 1 reveals the differences: \emph{annotators 5, 10, and 11} show similarly higher focus on the final frames (45-54), while \emph{annotator 12} focuses more on middle frames (30-35). They all predict the correct label ``happy'', but the different focus positions indicate that \emph{annotator 12}'s preference pattern for happiness in sample 1 differs from most annotators. Sample 2 shows: \emph{annotators 2, 8, and 13} exhibit similarly higher focus on the final frames (45-55), while \emph{annotator 12} focuses more on early frames (5-10). They all predict ``sad'' but \emph{annotator 12} demonstrates a different preference pattern.

\begin{figure*}
  \centering
  \begin{subfigure}{1.0\linewidth}
    \centering
    \includegraphics[width=0.85\linewidth]{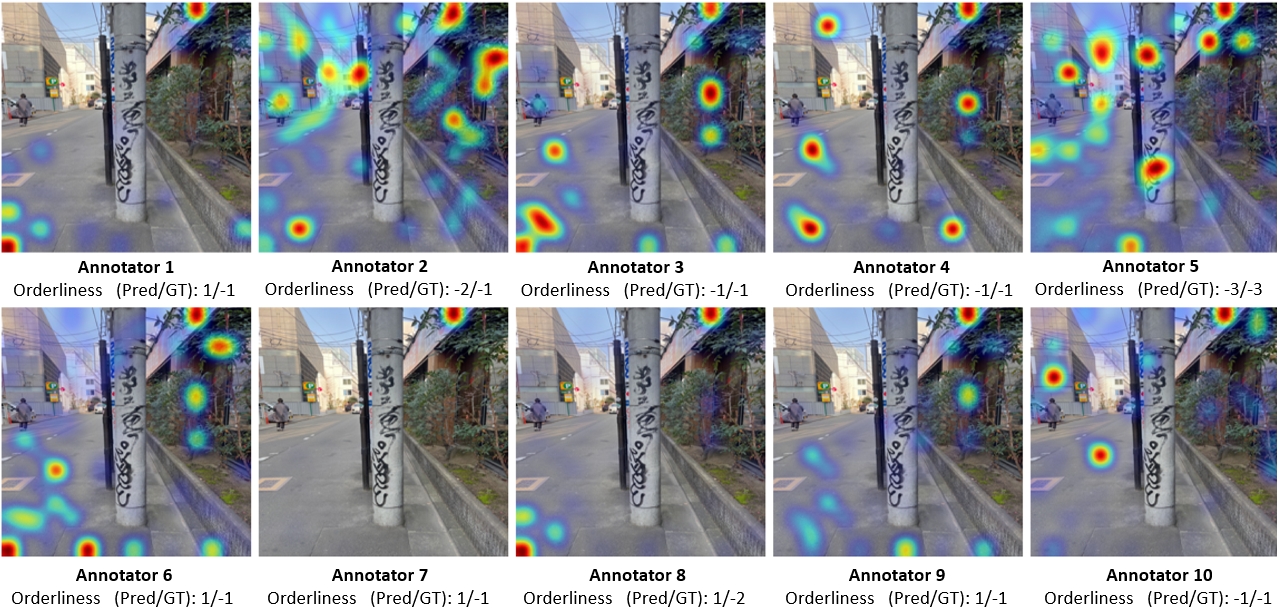}
    \caption{Sample 1 for orderliness perspective.}
    \label{fig:ia1}
  \end{subfigure}

  \vspace{2em}

  \begin{subfigure}{1.0\linewidth}
    \centering
    \includegraphics[width=0.85\linewidth]{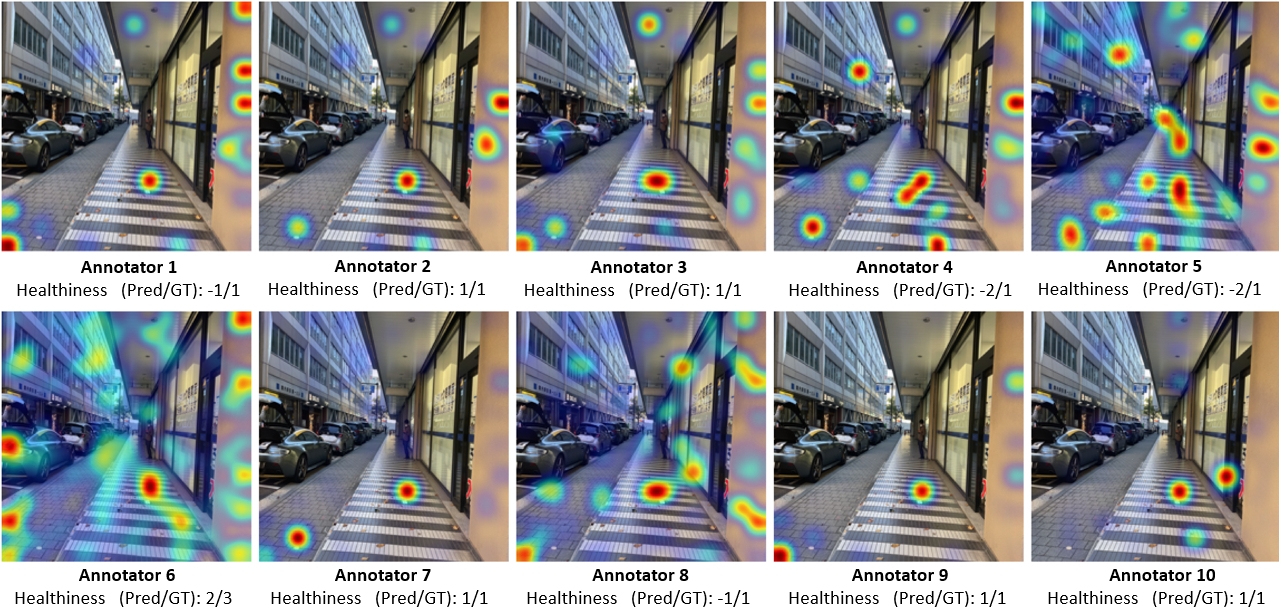}
    \caption{Sample 2 for healthiness perspective.}
    \label{fig:ia3}
  \end{subfigure}

  \caption{Additional sample experimental results visualize the annotator tendencies on the STREET image dataset. The tendencies of 10 annotators reveal distinct preferences.}
  \label{fig:ia}
\end{figure*}

As shown in Figure~\ref{fig:ia}, on the STREET dataset, annotators exhibit different tendencies through varying focus on different semantic elements within the same input image. In sample 1 from the orderliness perspective, the focus regions show differences: \emph{annotator 5} focuses more intensively on graffiti in the image compared with other annotators. The results show that prediction and ground truth are highly consistent in the emotion reflected in the semantics, corresponding to this annotator's expected lowest score. In sample 2 of the healthiness perspective, most annotators focus on uncleaned garbage and leaves, and only \emph{annotator 6} shows more focus on the surrounding environment, like cars and buildings, which might influence the result of not giving a low score.

\bibliographystyle{ACM-Reference-Format}
\bibliography{sample-base}

\end{document}